\documentclass[aps,prl,reprint,groupedaddress]{revtex4-1}
\usepackage{amsmath}
\usepackage{array}

\begin{document}

\title{SLOCC determinant invariants of order $2^{n/2}$ for even $n$ qubits}
\author{Xiangrong Li$^1$, Dafa Li$^2$}
%\date{\today }
\affiliation{$^1$ Department of Mathematics, University of
California, Irvine, CA 92697-3875, USA \\
$^2$ Department of mathematical sciences, Tsinghua
University, Beijing 100084 CHINA}

\begin{abstract}
% insert abstract here
In this paper, we study
SLOCC determinant invariants of order
$2^{n/2}$ for any even $n$ qubits
which satisfy the SLOCC determinant equations.
The determinant invariants can be constructed by a simple method
and the set of all these determinant invariants is complete 
with respect to permutations of qubits.
SLOCC entanglement classification can be achieved via the vanishing or not 
of the determinant invariants.
We exemplify the method for several even number of qubits,
with an emphasis on six qubits.
\end{abstract}

\maketitle

%\pacs{03.67.Mn}
% PACS Number: 03.67.Mn

% Use the \preprint command to place your local institutional report
% number in the upper righthand corner of the title page in preprint mode.
% Multiple \preprint commands are allowed.
% Use the 'preprintnumbers' class option to override journal defaults
% to display numbers if necessary
%\preprint{}

%Title of paper

% repeat the \author .. \affiliation  etc. as needed
% \email, \thanks, \homepage, \altaffiliation all apply to the current
% author. Explanatory text should go in the []'s, actual e-mail
% address or url should go in the {}'s for \email and \homepage.
% Please use the appropriate macro foreach each type of information

% \affiliation command applies to all authors since the last
% \affiliation command. The \affiliation command should follow the
% other information
% \affiliation can be followed by \email, \homepage, \thanks as well.

%\email[]{Your e-mail address}
%\homepage[]{Your web page}
%\thanks{}
%\altaffiliation{}

%Collaboration name if desired (requires use of superscriptaddress
%option in \documentclass). \noaffiliation is required (may also be
%used with the \author command).
%\collaboration can be followed by \email, \homepage, \thanks as well.
%\collaboration{}
%\noaffiliation

% insert suggested PACS numbers in braces on next line

% insert suggested keywords - APS authors don't need to do this
%\keywords{}

%\maketitle must follow title, authors, abstract, \pacs, and \keywords

% body of paper here - Use proper section commands
% References should be done using the \cite, \ref, and \label commands


Quantum entanglement is a key quantum mechanical resource in quantum
computation and information, such as quantum cryptography, quantum dense
coding and quantum teleportation \cite{Horodecki}. Whereas bipartite
entanglement has been well understood, multipartite entanglement remains
largely unexplored due to the exponential growth of complexity with the
number of qubits involved.

Functions in the coefficients of pure states which are invariant under
stochastic local operations and classical communication (SLOCC) play a vital
role in the study of entanglement classification
\cite{Dur,Miyake,Chterental,Cao,LDFPLA,LDF07a,LDF07b,LDFQIC09,Buniy,Viehmann}
as well as entanglement measures \cite{Luque,Osterloh05,LDF09b}.
Invariants for four
qubits have been presented in \cite{Luque}, and entanglement measures might
be built from the absolute values of these invariants. The three invariants
of order 4, denoted as $L$, $M$ and $N$, can be expressed in the form of
determinants. Invariants for five qubits have been highlighted in \cite%
{Luque06,Osterloh09}. To date, very few attempts have been made toward the
generalization to higher number of qubits. The SLOCC equations of degree 2
for even $n$ qubits and of degree 4 for odd $n$ qubits have been recently
established for two states equivalent under SLOCC \cite{LDF07a,LDF09b}. More
recently, for even $n$ qubits, the SLOCC determinant equations of degree $%
2^{n/2}$ has been established and four determinant invariants of order $%
2^{n/2}$ have been obtained \cite{LDFJPA}. In light of the SLOCC determinant
equations, several different genuine entangled states of even $n$ qubits
inequivalent to the $|GHZ\rangle$, $|W\rangle$, and Dicke states have been
constructed.

In this paper, we construct $\binom{n-1}{n/2-1}$ SLOCC determinant
invariants of order $2^{n/2}$ for any even $n$ qubits which satisfy the
SLOCC determinant equations. We also demonstrate the completeness of the set
of all these determinant invariants with respect to permutations of qubits. 
For six qubits, we explicitly derive all the ten SLOCC
determinants of order $8$. The determinant invariants can be used for SLOCC
classification of any even $n$ qubits. Finally, we illustrate the
application of the equations and invariants by proposing a genuine entangled
state of even $n$ qubits and showing that it is inequivalent to the $%
|GHZ\rangle $, $|W\rangle $, and Dicke states.

We write the state $|\psi ^{\prime }\rangle $ of even $n$ qubits as $|\psi
^{\prime }\rangle =\sum_{i=0}^{2^{n}-1}a_{i}|i\rangle $. We associate to the
state $|\psi ^{\prime }\rangle$ a $2^{n/2}$ by $2^{n/2}$ coefficient matrix $%
M(a,n)$ whose entries are the coefficients $a_{0}, a_{1}, \cdots,
a_{2^{n}-1} $ arranged in ascending lexicographical order. To illustrate, we
list $M(a,4) $ below as:
\begin{equation}
M(a,4)=\left(
\begin{array}{cccc}
a_{0} & a_{1} & a_{2} & a_{3} \\
a_{4} & a_{5} & a_{6} & a_{7} \\
a_{8} & a_{9} & a_{10} & a_{11} \\
a_{12} & a_{13} & a_{14} & a_{15}%
\end{array}
\right).
\end{equation}

Let the state $|\psi \rangle $\ of even $n$ qubits be $|\psi \rangle
=\sum_{i=0}^{2^{n}-1}b_{i}|i\rangle $. It is well known that the states $%
|\psi \rangle $ and $|\psi ^{\prime }\rangle $ are equivalent under SLOCC if
and only if there exist local invertible operators $\mathcal{A}_{1}$, $%
\mathcal{A}_{2},\cdots ,\mathcal{A}_{n}$ such that \cite{Dur}
\begin{equation}
|\psi ^{\prime }\rangle =\underbrace{\mathcal{A}_{1}\otimes \mathcal{A}%
_{2}\otimes \cdots \otimes \mathcal{A}_{n}}_{n}|\psi \rangle .  \label{eq_1}
\end{equation}

Let $M(b,n)$ be obtained from $M(a,n)$ by replacing $a$ by $b$ . Then the
following SLOCC determinant equation holds \cite{LDFJPA}:
\begin{align}
& \det M(a,n)  \notag \\
=& \det M(b,n)[\det (\mathcal{A}_{1})\cdots \det (\mathcal{A}%
_{n})]^{2^{(n-2)/2}}.  \label{det-1}
\end{align}%
We refer to any determinant that satisfies Eq. (\ref{det-1}) as a SLOCC
determinant invariant of order $2^{n/2}$ for even $n$ qubits. In particular,
$\det M(a,n)$ is such a determinant invariant. Several other determinant
invariants have recently been obtained in \cite{LDFJPA}. The aim of this
paper is to construct all the determinant invariants satisfying
Eq. (\ref{det-1}).

We write $|\psi ^{\prime }\rangle $ in terms of an orthogonal basis as $%
|\psi ^{\prime }\rangle =\sum a_{i_{1}i_{2}\cdots i_{n}}|i_{1}i_{2}$ $\cdots
i_{n}\rangle $, where $i_1i_2\cdots i_n$ is the $n$-bit binary form of the
index $i$. Inspection of the structure of the matrix $M(a,n)$ reveals that
the coefficient $a_{i_{1}\cdots i_{n/2}i_{n/2+1}\cdots i_{n}}$ of the state $%
|\psi ^{\prime }\rangle $ is the entry in the $(i_{1}\cdots i_{n/2})$th row
and $(i_{n/2+1}$$\cdots$$i_{n})$th column of the matrix (define the topmost
row as the $0$th row and the leftmost column as the $0$th column). In other
words, bits $1,\cdots,{n/2}$ specify the row number, and the rest bits
specify the column number. We observe that using different bits to specify
the row number might result in matrices whose determinants are different
from that of $M(a,n)$. Since $n/2$ bits are needed to specify the row number
for square matrices, this amounts to $\binom{n}{n/2}$ different ways. But,
as can easily be verified, exchanging the row and column bits of a matrix is
equivalent to transposing the matrix. This gives a total of $\frac{1}{2}%
\binom{n}{n/2}=\binom{n-1}{n/2-1} \label{eq_total}$ different determinants.
As will be seen later, these determinants satisfy the SLOCC determinant
equations and form a complete set of determinant invariants of order $%
2^{n/2} $ of even $n$ qubits with respect to permutations of qubits. 
We can construct the determinants in the
following way: use bit $n/2$ together with $n/2-1$ other bits selected from
the rest $n-1$ bits to specify the row number and the remaining $n/2$ bits
to specify the column number. We exemplify this for $n=4$. Using bits 1 and
2 to specify the row number yields
\begin{align}
D_{4}^{1}=\left\vert
\begin{array}{*{4}{>{\centering\arraybackslash$}p{0.5cm}<{$}}}
a_{0} & a_{1} & a_{2} & a_{3} \\
a_{4} & a_{5} & a_{6} & a_{7} \\
a_{8} & a_{9} & a_{10} & a_{11} \\
a_{12} & a_{13} & a_{14} & a_{15}%
\end{array}%
\right\vert. \label{eq_d41}
\end{align}%
Using bits 2 and 3 to specify the row number yields
\begin{align}
D_{4}^{2}=\left\vert
\begin{array}{*{4}{>{\centering\arraybackslash$}p{0.5cm}<{$}}}
a_{0} & a_{1} & a_{8} & a_{9} \\
a_{2} & a_{3} & a_{10} & a_{11} \\
a_{4} & a_{5} & a_{12} & a_{13} \\
a_{6} & a_{7} & a_{14} & a_{15}%
\end{array}%
\right\vert. \label{eq_d42}
\end{align}%
Using bits 2 and 4 to specify the row number yields
\begin{align}
D_{4}^{3}=\left\vert
\begin{array}{*{4}{>{\centering\arraybackslash$}p{0.5cm}<{$}}}
a_{0} & a_{2} & a_{8} & a_{10} \\
a_{1} & a_{3} & a_{9} & a_{11} \\
a_{4} & a_{6} & a_{12} & a_{14} \\
a_{5} & a_{7} & a_{13} & a_{15}%
\end{array}%
\right\vert. \label{eq_d43}
\end{align}%
Each of the above three determinants can be verified to satisfy Eq. (\ref%
{det-1}) by solving Eq. (\ref{eq_1}) for the coefficients of $|\psi ^{\prime
}\rangle $ and then substituting the coefficients into the corresponding
Eqs. (\ref{eq_d41})-(\ref{eq_d43}), thereby revealing that all the three
determinants are SLOCC determinant invariants of order 4 for four qubits
\cite{LDFJPA}. It is worth noting that the above three determinants,
ignoring the sign, turn out to be same as the ones given in \cite{Luque}.

While the determinants of order 4 for four qubits can be verified to satisfy
Eq. (\ref{det-1}), this is an extremely difficult task for large number of
qubits. To solve this problem, we will resort to permutations. Suppose that
we use bit $n/2$ together with $n/2-1$ other bits $\ell_{1}$, $\ell_{2},\cdots
,\ell_{n/2-1}$ selected from the rest $n-1$ bits to specify the row number and
the remaining $n/2$ bits to specify the column number. This gives a
determinant of order $2^{n/2}$. We will show that this determinant can be
obtained by applying a permutation to $\det M(a,n)$. This can be seen as
follows. Let
\begin{align}
C=\{\ell_{1}, \ell_{2},\cdots ,\ell_{n/2-1}\}\cap \{1,2,\cdots,n/2-1\},
\end{align}
i.e. $C$ consists of those among the first $n/2-1$ bits which are used to
specify the row number. Consider the following two sets of bits:
\begin{align}
\{t_{1},t_{2},\cdots,t_{k}\}&=\{\ell_{1},\ell_{2},\cdots ,\ell_{n/2-1}\}/C, \\
\{r_{1},r_{2},\cdots,r_{k}\}&=\{1,2,\cdots ,n/2-1\}/C,
\end{align}
for some $0\leq k\leq n/2-1$. Here $r_1,\cdots,r_k$ are those among the
first $n/2$ bits which are used to specify the column number, and $%
t_1,\cdots,t_k$ are those among the last $n/2$ bits which are used to
specify the row number. Define the permutation
\begin{equation}
\sigma =(r_{1},t_{1})(r_{2},t_{2})\cdots (r_{k},t_{k}).  \label{permut}
\end{equation}%
If $k=0$, we define $\sigma=I$. It is trivial to see that, ignoring the
sign, the determinant constructed above is equal to $\sigma \det M(a,n)$. To
find all the determinants of order $2^{n/2}$, we can simply exhaust all
possible values of $r_1,\cdots,r_k$, $t_1,\cdots,t_k$, and $k$, i.e. for all
$1\leq r_{1}<r_{2}<\cdots <r_{k}\leq n/2-1$, $n/2<t_{1}<t_{2}<\cdots
<t_{k}\leq n$, and $k$ varies from 0 to $n/2-1$. Inspection of the above
condition yields $\sum_{k=0}^{n/2-1}{\binom{n/2-1}{k}}{\binom{n/2}{k}}$
\noindent $={\binom{n-1}{n/2-1}}$ different permutations $\sigma$, which
give rise to equally as many different determinants $\sigma\det M(a,n)$ of
order $2^{n/2}$.

Now, simply taking the permutations $\sigma $ to both sides of Eq. (\ref%
{det-1}) yield the following determinant equations:
\begin{align}
& \sigma \det M(a,n)  \notag \\
=& \sigma \det M(b,n)[\det (\mathcal{A}_{1})\cdots \det (\mathcal{A}%
_{n})]^{2^{(n-2)/2}}.  \label{main-eq1}
\end{align}
It follows immediately from Eq. (\ref{main-eq1}) that $\sigma \det M(a,n)$
are determinant invariants of order $2^{n/2}$.

For the sake of completeness, we first do a simple manipulation of Eq. (\ref%
{permut}). If we make use of the fact that
\begin{align}
(r_k,t_k)=(1,t_k)(1,r_k)(1,t_k),  \label{transp}
\end{align}
we are led to the following equation (ignoring the sign):
\begin{align}
(r_{i},t_{i})\det M(a,n)=(1,r_{i})(1,t_{i})\det M(a,n).  \label{eq_sigma}
\end{align}
That Eq. (\ref{eq_sigma}) holds is easily confirmed upon realizing that to
take transposition $(1,t_{i})$ to the determinant $(1,r_{i})(1,t_{i})\det
M(a,n)$ is equivalent to interchanging two rows of the determinant. This
leads to the following expression for $\sigma$:
\begin{align}
\sigma =(1,r_{1})(1,t_{1})(1,r_{2})(1,t_{2})\cdots (1,r_{k})(1,t_{k}).
\label{permut-2}
\end{align}%
Indeed, Eq. (\ref{permut-2}) is usually more convenient to use than Eq. (\ref%
{permut}). It can be demonstrated that applying a transposition in the form $%
(1,i)$ with $i=1,\cdots,n$ to the set formed by the determinant invariants
constructed above always yields the same set (ignoring the sign). Since any
permutation can be expressed as a product of transpositions in the form $%
(1,i)$, this demonstrates the completeness of the set of all these determinant
invariants. This will be illustrated in discussing the cases 
$n=4$ and $n=6$ below.

We now proceed to present the determinant invariants for several even number
of qubits.

$n=2$: for two qubits, there is only one determinant invariant $\left\vert
\begin{array}{cc}
a_{0} & a_{1} \\
a_{2} & a_{3}%
\end{array}%
\right\vert$ of order 2 (see also \cite{LDF07a}).

$n=4$: for four qubits, there are three determinant invariants of order 4,
namely, $D_{4}^{1}$, $D_{4}^{2}$, and $D_{4}^{3}$ 
(see Eqs. (\ref{eq_d41})-(\ref{eq_d43})). 
Note that $D_4^2=(1,3)D_4^1$ and $D_4^3=(1,4)D_4^1$.

We now argue that the above three determinants form a complete set of
determinant invariants of order 4 for four qubits with respect to permutations 
of qubits. This can be seen as follows. 
Applying any transposition $(1,i)$ with $i=1,\cdots, 4$ to any one
of the three determinant invariants always yields a determinant invariant in
the same set. This demonstrates the completeness of the set formed by these
determinant invariants. The results are summarized in table \ref{table1}.
\begin{table}[tbph]
\caption{Completeness of the determinant invariants for four qubits 
with respect to permutations of qubits}
\label{table1}\renewcommand\arraystretch{1.5} \center
\begin{ruledtabular}
\begin{tabular}{cccc}
trans & \multicolumn{2}{c}{\quad \quad \quad \quad determinant invariants} & \\\hline
$(1,1)$ & $D_4^1$ & $D_4^2$ & $D_4^3$ \\
$(1,2)$ & $D_4^1$ & $D_4^3$ & $D_4^2$ \\
$(1,3)$ & $D_4^2$ & $D_4^1$ & $D_4^3$ \\
$(1,4)$ & $D_4^3$ & $D_4^2$ & $D_4^1$ \\
\end{tabular}
\end{ruledtabular}
\end{table}

\begin{table*}[tbp]
\caption{Ten determinant invariants for six qubits}
\label{table2}\renewcommand\arraystretch{1.5}
\begin{ruledtabular}
\begin{tabular}{ccccccccccc}
$\sigma$ & $I$ & $(1,4)$ & $(1,5)$ & $(1,6)$ & $(1,2)(1,4)$ & $(1,2)(1,5)$
& $%
(1,2)(1,6) $ & $(1,4)(1,2)(1,5)^a$ & $(1,4)(1,2)(1,6)^b$
& $(1,5)(1,2)(1,6)^c$ \\
$\det$ & $D_6^1$ & $D_6^2 $ & $D_6^3 $ & $D_6^4 $ & $D_6^5$ & $D_6^6$
& $D_6^7$ & $D_6^8$ & $D_6^9$ & $D_6^{10}$ \\
\end{tabular}
\end{ruledtabular}
$^a$ $(1,4)(1,2)(1,5)D_6^1 =(1,3)(1,6)D_6^1=D_{6}^{8}$.\hfill\hfill

$^b$ $(1,4)(1,2)(1,6)D_6^1=(1,3)(1,5)D_6^1 =D_{6}^{9}$.\hfill\hfill

$^c$ $(1,5)(1,2)(1,6)D_6^1=(1,3)(1,4)D_6^1=D_{6}^{10}$.\hfill\hfill
\end{table*}

$n=6$: for six qubits, there are ten determinant invariants of order 8. In
table \ref{table2}, we list the permutations $\sigma$ and the corresponding
determinant invariants by virtue of Eq. (\ref{permut-2}). The determinant
invariants are explicitly given as follows:
\begin{equation*}
D_{6}^{1}=\left\vert
\begin{array}{cccccccc}
a_{0} & a_{1} & a_{2} & a_{3} & a_{4} & a_{5} & a_{6} & a_{7} \\
a_{8} & a_{9} & a_{10} & a_{11} & a_{12} & a_{13} & a_{14} & a_{15} \\
a_{16} & a_{17} & a_{18} & a_{19} & a_{20} & a_{21} & a_{22} & a_{23} \\
a_{24} & a_{25} & a_{26} & a_{27} & a_{28} & a_{29} & a_{30} & a_{31} \\
a_{32} & a_{33} & a_{34} & a_{35} & a_{36} & a_{37} & a_{38} & a_{39} \\
a_{40} & a_{41} & a_{42} & a_{43} & a_{44} & a_{45} & a_{46} & a_{47} \\
a_{48} & a_{49} & a_{50} & a_{51} & a_{52} & a_{53} & a_{54} & a_{55} \\
a_{56} & a_{57} & a_{58} & a_{59} & a_{60} & a_{61} & a_{62} & a_{63}%
\end{array}%
\right\vert
\end{equation*}%
\begin{equation*}
D_{6}^{2}=\left\vert
\begin{array}{cccccccc}
a_{0} & a_{1} & a_{2} & a_{3} & a_{32} & a_{33} & a_{34} & a_{35} \\
a_{4} & a_{5} & a_{6} & a_{7} & a_{36} & a_{37} & a_{38} & a_{39} \\
a_{8} & a_{9} & a_{10} & a_{11} & a_{40} & a_{41} & a_{42} & a_{43} \\
a_{12} & a_{13} & a_{14} & a_{15} & a_{44} & a_{45} & a_{46} & a_{47} \\
a_{16} & a_{17} & a_{18} & a_{19} & a_{48} & a_{49} & a_{50} & a_{51} \\
a_{20} & a_{21} & a_{22} & a_{23} & a_{52} & a_{53} & a_{54} & a_{55} \\
a_{24} & a_{25} & a_{26} & a_{27} & a_{56} & a_{57} & a_{58} & a_{59} \\
a_{28} & a_{29} & a_{30} & a_{31} & a_{60} & a_{61} & a_{62} & a_{63}%
\end{array}%
\right\vert
\end{equation*}%
\begin{equation*}
D_{6}^{3}=\left\vert
\begin{array}{cccccccc}
a_{0} & a_{1} & a_{4} & a_{5} & a_{32} & a_{33} & a_{36} & a_{37} \\
a_{2} & a_{3} & a_{6} & a_{7} & a_{34} & a_{35} & a_{38} & a_{39} \\
a_{8} & a_{9} & a_{12} & a_{13} & a_{40} & a_{41} & a_{44} & a_{45} \\
a_{10} & a_{11} & a_{14} & a_{15} & a_{42} & a_{43} & a_{46} & a_{47} \\
a_{16} & a_{17} & a_{20} & a_{21} & a_{48} & a_{49} & a_{52} & a_{53} \\
a_{18} & a_{19} & a_{22} & a_{23} & a_{50} & a_{51} & a_{54} & a_{55} \\
a_{24} & a_{25} & a_{28} & a_{29} & a_{56} & a_{57} & a_{60} & a_{61} \\
a_{26} & a_{27} & a_{30} & a_{31} & a_{58} & a_{59} & a_{62} & a_{63}%
\end{array}%
\right\vert
\end{equation*}%
\begin{equation*}
D_{6}^{4}=\left\vert
\begin{array}{cccccccc}
a_{0} & a_{2} & a_{4} & a_{6} & a_{32} & a_{34} & a_{36} & a_{38} \\
a_{8} & a_{10} & a_{12} & a_{14} & a_{40} & a_{42} & a_{44} & a_{46} \\
a_{1} & a_{3} & a_{5} & a_{7} & a_{33} & a_{35} & a_{37} & a_{39} \\
a_{9} & a_{11} & a_{13} & a_{15} & a_{41} & a_{43} & a_{45} & a_{47} \\
a_{16} & a_{18} & a_{20} & a_{22} & a_{48} & a_{50} & a_{52} & a_{54} \\
a_{24} & a_{26} & a_{28} & a_{30} & a_{56} & a_{58} & a_{60} & a_{62} \\
a_{17} & a_{19} & a_{21} & a_{23} & a_{49} & a_{51} & a_{53} & a_{55} \\
a_{25} & a_{27} & a_{29} & a_{31} & a_{57} & a_{59} & a_{61} & a_{63}%
\end{array}%
\right\vert
\end{equation*}%
\begin{equation*}
D_{6}^{5}=\left\vert
\begin{array}{cccccccc}
a_{0} & a_{1} & a_{2} & a_{3} & a_{16} & a_{17} & a_{18} & a_{19} \\
a_{4} & a_{5} & a_{6} & a_{7} & a_{20} & a_{21} & a_{22} & a_{23} \\
a_{8} & a_{9} & a_{10} & a_{11} & a_{24} & a_{25} & a_{26} & a_{27} \\
a_{12} & a_{13} & a_{14} & a_{15} & a_{28} & a_{29} & a_{30} & a_{31} \\
a_{32} & a_{33} & a_{34} & a_{35} & a_{48} & a_{49} & a_{50} & a_{51} \\
a_{36} & a_{37} & a_{38} & a_{39} & a_{52} & a_{53} & a_{54} & a_{55} \\
a_{40} & a_{41} & a_{42} & a_{43} & a_{56} & a_{57} & a_{58} & a_{59} \\
a_{44} & a_{45} & a_{46} & a_{47} & a_{60} & a_{61} & a_{62} & a_{63}%
\end{array}%
\right\vert
\end{equation*}%
\begin{equation*}
D_{6}^{6}=\left\vert
\begin{array}{cccccccc}
a_{0} & a_{2} & a_{8} & a_{10} & a_{32} & a_{34} & a_{40} & a_{42} \\
a_{1} & a_{3} & a_{9} & a_{11} & a_{33} & a_{35} & a_{41} & a_{43} \\
a_{4} & a_{6} & a_{12} & a_{14} & a_{36} & a_{38} & a_{44} & a_{46} \\
a_{5} & a_{7} & a_{13} & a_{15} & a_{37} & a_{39} & a_{45} & a_{47} \\
a_{16} & a_{18} & a_{24} & a_{26} & a_{48} & a_{50} & a_{56} & a_{58} \\
a_{17} & a_{19} & a_{25} & a_{27} & a_{49} & a_{51} & a_{57} & a_{59} \\
a_{20} & a_{22} & a_{28} & a_{30} & a_{52} & a_{54} & a_{60} & a_{62} \\
a_{21} & a_{23} & a_{29} & a_{31} & a_{53} & a_{55} & a_{61} & a_{63}%
\end{array}%
\right\vert
\end{equation*}%
\begin{equation*}
D_{6}^{7}=\left\vert
\begin{array}{cccccccc}
a_{0} & a_{2} & a_{4} & a_{6} & a_{16} & a_{18} & a_{20} & a_{22} \\
a_{1} & a_{3} & a_{5} & a_{7} & a_{17} & a_{19} & a_{21} & a_{23} \\
a_{8} & a_{10} & a_{12} & a_{14} & a_{24} & a_{26} & a_{28} & a_{30} \\
a_{9} & a_{11} & a_{13} & a_{15} & a_{25} & a_{27} & a_{29} & a_{31} \\
a_{32} & a_{34} & a_{36} & a_{38} & a_{48} & a_{50} & a_{52} & a_{54} \\
a_{33} & a_{35} & a_{37} & a_{39} & a_{49} & a_{51} & a_{53} & a_{55} \\
a_{40} & a_{42} & a_{44} & a_{46} & a_{56} & a_{58} & a_{60} & a_{62} \\
a_{41} & a_{43} & a_{45} & a_{47} & a_{57} & a_{59} & a_{61} & a_{63}%
\end{array}%
\right\vert
\end{equation*}%
\begin{equation*}
D_{6}^{8}=\left\vert
\begin{array}{cccccccc}
a_{0} & a_{2} & a_{4} & a_{6} & a_{8} & a_{10} & a_{12} & a_{14} \\
a_{1} & a_{3} & a_{5} & a_{7} & a_{9} & a_{11} & a_{13} & a_{15} \\
a_{16} & a_{18} & a_{20} & a_{22} & a_{24} & a_{26} & a_{28} & a_{30} \\
a_{17} & a_{19} & a_{21} & a_{23} & a_{25} & a_{27} & a_{29} & a_{31} \\
a_{32} & a_{34} & a_{36} & a_{38} & a_{40} & a_{42} & a_{44} & a_{46} \\
a_{33} & a_{35} & a_{37} & a_{39} & a_{41} & a_{43} & a_{45} & a_{47} \\
a_{48} & a_{50} & a_{52} & a_{54} & a_{56} & a_{58} & a_{60} & a_{62} \\
a_{49} & a_{51} & a_{53} & a_{55} & a_{57} & a_{59} & a_{61} & a_{63}%
\end{array}%
\right\vert
\end{equation*}
\begin{equation*}
D_{6}^{9}=\left\vert
\begin{array}{cccccccc}
a_{0} & a_{1} & a_{4} & a_{5} & a_{8} & a_{9} & a_{12} & a_{13} \\
a_{2} & a_{3} & a_{6} & a_{7} & a_{10} & a_{11} & a_{14} & a_{15} \\
a_{16} & a_{17} & a_{20} & a_{21} & a_{24} & a_{25} & a_{28} & a_{29} \\
a_{18} & a_{19} & a_{22} & a_{23} & a_{26} & a_{27} & a_{30} & a_{31} \\
a_{32} & a_{33} & a_{36} & a_{37} & a_{40} & a_{41} & a_{44} & a_{45} \\
a_{34} & a_{35} & a_{38} & a_{39} & a_{42} & a_{43} & a_{46} & a_{47} \\
a_{48} & a_{49} & a_{52} & a_{53} & a_{56} & a_{57} & a_{60} & a_{61} \\
a_{50} & a_{51} & a_{54} & a_{55} & a_{58} & a_{59} & a_{62} & a_{63}%
\end{array}%
\right\vert
\end{equation*}

\begin{equation*}
D_{6}^{10}=\left\vert
\begin{array}{cccccccc}
a_{0} & a_{1} & a_{2} & a_{3} & a_{8} & a_{9} & a_{10} & a_{11} \\
a_{4} & a_{5} & a_{6} & a_{7} & a_{12} & a_{13} & a_{14} & a_{15} \\
a_{16} & a_{17} & a_{18} & a_{19} & a_{24} & a_{25} & a_{26} & a_{27} \\
a_{20} & a_{21} & a_{22} & a_{23} & a_{28} & a_{29} & a_{30} & a_{31} \\
a_{32} & a_{33} & a_{34} & a_{35} & a_{40} & a_{41} & a_{42} & a_{43} \\
a_{36} & a_{37} & a_{38} & a_{39} & a_{44} & a_{45} & a_{46} & a_{47} \\
a_{48} & a_{49} & a_{50} & a_{51} & a_{56} & a_{57} & a_{58} & a_{59} \\
a_{52} & a_{53} & a_{54} & a_{55} & a_{60} & a_{61} & a_{62} & a_{63}%
\end{array}%
\right\vert
\end{equation*}

An argument analogous to the one for $n=4$ establishes the completeness of 
the set formed by the ten determinant invariants. 
The results are summarized in table \ref{table3}.

\begin{table}[tbph]
\caption{Completeness of the determinant invariants for six qubits
with respect to permutations of qubits}
\label{table3}\renewcommand\arraystretch{1.5} \center
\begin{ruledtabular}
\begin{tabular}{ccccccccccc}
trans & \multicolumn{10}{c}{determinant invariants} \\ \hline
$(1,1)$ & $D_6^1$ & $D_6^2$ & $D_6^3$ & $D_6^4$ & $D_6^5$ & $D_6^6$ &
$D_6^7$ & $D_6^8$ & $D_6^9$ & $D_6^{10}$ \\
$(1,2)$ & $D_6^1$ & $D_6^5$ & $D_6^6$ & $D_6^7$ & $D_6^2$ & $D_6^3$ &
$D_6^4$ & $D_6^8$ & $D_6^9$ & $D_6^{10}$ \\
$(1,3)$ & $D_6^1$ & $D_6^{10}$ & $D_6^9$ & $D_6^8$ & $D_6^5$ & $D_6^6$ &
$D_6^7$ & $D_6^4$ & $D_6^3$ & $D_6^2$ \\
$(1,4)$ & $D_6^2$ & $D_6^1$ & $D_6^3$ & $D_6^4$ & $D_6^5$ & $D_6^8$ &
$D_6^9$ & $D_6^6$ & $D_6^7$ & $D_6^{10}$ \\
$(1,5)$ & $D_6^3$ & $D_6^2$ & $D_6^1$ & $D_6^4$ & $D_6^8$ & $D_6^6$ &
$D_6^{10}$ & $D_6^5$ & $D_6^9$ & $D_6^7$ \\
$(1,6)$ & $D_6^4$ & $D_6^2$ & $D_6^3$ & $D_6^1$ & $D_6^9$ & $D_6^{10}$ &
$D_6^7$ & $D_6^8$ & $D_6^5$ & $D_6^6$ \\
\end{tabular}
\end{ruledtabular}
\end{table}

$n=8$: for eight qubits, there are 35 determinant invariants of order 16
(not presented here due to space limitations).

Finally, it follows from Eq. (\ref{main-eq1}) that if two $n$-qubit states $%
|\psi ^{\prime }\rangle $ and $|\psi \rangle $ are SLOCC equivalent, then $%
\sigma \det M(a,n)$ vanishes if and only if $\sigma \det M(b,n)$ vanishes.
On the other hand, if one of the determinants $\sigma \det M(a,n)$ and $%
\sigma \det M(b,n)$ vanishes while the other does not, then the states $%
|\psi ^{\prime }\rangle $ and $|\psi \rangle $ belong to different SLOCC
equivalent classes. Thus, each determinant divides the space of the pure
states of even $n$ qubits into two inequivalent subspaces under SLOCC. Let $%
c=\binom{n-1}{n/2-1}$. In total, $c$ determinants divide the space into $%
2^{c}$ subspaces (or families) under SLOCC. Here each family $F_{\delta
_{1}\cdots \delta _{c}}$ is defined as $F_{\delta _{1}\cdots \delta
_{c}}=S_{\delta _{1}}^{(1)}\cap \cdots \cap S_{\delta _{c}}^{(c)}$, where $%
\delta _{i}=0$ or $1$, $S_{0}^{(i)}=\{|\psi \rangle |D_{n}^{i}=0\}$, and $%
S_{1}^{(i)}=\{|\psi \rangle |D_{n}^{i}\neq 0\}$. Clearly, some 
families include infinite SLOCC classes. 
It is straightforward to see that if two states are SLOCC equivalent then they 
belong to the same family. However, the converse does not hold, i.e.
two states belonging to the same family are not necessarily SLOCC equivalent.

As an application of the SLOCC determinant equations and invariants,
consider, for example, the following genuine entangled state for six qubits:
\begin{equation}
|\chi \rangle =(1/\sqrt{8})(|0\rangle +|5\rangle +|18\rangle +|23\rangle
+|40\rangle +|45\rangle +|58\rangle -|63\rangle ).  \notag
\end{equation}
We observe that all the non-zero coefficients of $|\chi \rangle$ lie on the
diagonal of $D_6^{10}$. This leads to non-vanishing $D_6^{10}$ for $|\chi
\rangle $. However, $D_6^{10}$ vanishes for the $|GHZ\rangle $, $|W\rangle $
and Dicke states. In light of Eq. (\ref{main-eq1}), $|\chi \rangle$ is
inequivalent to the $|GHZ\rangle $, $|W\rangle $ and Dicke states under
SLOCC. For more examples, see \cite{LDFJPA}. 

In summary, we have constructed the set of all determinant invariants
of order $2^{n/2}$ for any even $n$ qubits and showed that the set is
complete with respect to permutations of qubits. 
We have presented a simple formula for constructing
the determinant invariants and given several examples for even $n$. The
determinant invariants can be used for SLOCC classification of any even $n$
qubits and the absolute values of the determinant invariants 
can be considered as entanglement measures.
Finally, a more fundamental problem is whether the determinant 
invariants are independent.

\begin{acknowledgments}
The paper was supported by NSFC (Grant No.10875061)
and Tsinghua National Laboratory for Information Science and Technology.
\end{acknowledgments}

\end{document}